\documentclass[11pt]{article}
\usepackage{amsfonts}
\usepackage{amsmath}
\usepackage{amssymb}
\usepackage{amsthm}
\usepackage{algorithm}
\usepackage{algorithmic}
\usepackage{float}
\usepackage{graphics}
\usepackage{graphicx}
\usepackage{geometry}
\usepackage{enumitem}
\usepackage{indentfirst}
\usepackage[utf8]{inputenc}
\usepackage{float}
\usepackage{caption}
\usepackage{subcaption}
\usepackage[labelfont=bf]{caption}
\newtheorem{lemma}{Lemma}
\newtheorem{theorem}{Theorem}
\newtheorem{definition}{Definition}
\usepackage{hyperref}
\usepackage{fancyhdr}

\title{Revisiting Garg's 2-Approximation Algorithm\\ for the $k$-MST Problem in Graphs}
\author{Emmett Breen \and Renee Mirka \and Zichen Wang \and David P. Williamson}
\date{\{ejb284, rem379, zw336, davidpwilliamson\}@cornell.edu\\ Cornell University, Ithaca, NY 14850}

\begin{document}

\maketitle

\fancyfoot[R]{\scriptsize{Copyright \textcopyright\ 2023 by SIAM\\
Unauthorized reproduction of this article is prohibited}}

\begin{abstract}
    This paper revisits the 2-approximation algorithm for $k$-MST presented by Garg \cite{Garg05} in light of a recent paper of Paul et al.\ \cite{PaulFFSW20}. In the $k$-MST problem, the goal is to return a tree spanning $k$ vertices of minimum total edge cost. Paul et al.\ \cite{PaulFFSW20} extend Garg's primal-dual subroutine to improve the approximation ratios for the budgeted prize-collecting traveling salesman and minimum spanning tree problems. We follow their algorithm and analysis to provide a cleaner version of Garg's result. Additionally, we introduce the novel concept of a kernel which allows an easier visualization of the stages of the algorithm and a clearer understanding of the pruning phase. Other notable updates include presenting a linear programming formulation of the $k$-MST problem, including pseudocode, replacing the coloring scheme used by Garg with the simpler concept of neutral sets, and providing an explicit potential function.
\end{abstract}


\section{Introduction}
Given as input an undirected graph $G=(V,E)$ and non-negative costs $c_e$ for each edge $e\in E$, the goal of the $k$-MST problem is to find a tree spanning at least $k$ vertices of minimum total cost. In the rooted case of the problem, a special root vertex $r$ is denoted which must be contained in the final tree, whereas no such $r$ exists for the unrooted case. These two cases are equivalent for the $k$-MST problem: Garg \cite{Garg05} shows that an $\alpha$-approximation algorithm for one version can be translated into an $\alpha$-approximation algorithm for the other. An $\alpha$-approximation algorithm for $k$-MST is a polynomial-time algorithm which returns a solution at most $\alpha$ times the cost of the optimal tree. 

Approximation algorithms have been considered for $k$-MST since Ravi et al.\ introduced the problem \cite{RaviSMRR94}.  They proved the problem is NP-hard and presented a $3\sqrt{k}$-approximation algorithm. Since then, one line of work by numerous authors \cite{AABV1995,BRV96, Garg1996, ARYA1998117} developed a $(2 + \epsilon)$-approximation algorithm running in time polynomial in $\epsilon^{-1}$ \cite{AroraK06}. Currently, the best known approximation is a 2-approximation algorithm by Garg using primal-dual techniques \cite{Garg05}. A separate line of work considers a special case of $k$-MST where the input is given by $n$ points in the plane with edge costs determined by the Euclidean metric \cite{RaviSMRR94,GH94,Eppstein1997,BCV1995,MBCV98, Arora98,Mitchell99}.

This paper focuses on Garg's result. His 2-approximation algorithm is not only the best known for $k$-MST, but Garg also shows that his algorithm leads to a 2-approximation algorithm for the $k$-TSP problem (where one seeks a min-cost tour on at least $k$ vertices instead of a tree and edge costs satisfy the triangle inequality) and 3-approximation algorithm for the budgeted version of $k$-MST (where a budget $B$ is given and one seeks to maximize the number of vertices in a tree of cost at most $B$) due to an observation by Johnson et al.\ \cite{JMP2000}. More recently, Paul et al.\ have extended Garg's technique to improve the approximation algorithms for the budgeted prize-collecting traveling salesman and minimum spanning tree problems \cite{PaulFFSW20}. We revisit Garg's algorithm and analysis in light of the paper of Paul et al.  In particular, we seek to provide a cleaner version of Garg's algorithm and analysis by adapting the results of Paul et al. We use Paul et al.'s explicit potential function and use of \textit{neutral} sets to replace the coloring scheme used by Garg. Both of these changes improve the clarity and comprehensibility of the initial analysis. Additionally, and distinct from both previous papers, we introduce the notion of a \textit{kernel} in this primal-dual algorithm; its use in part enables a clearer visualization of the mechanics of the algorithm. We believe this perspective yields a more accessible version of Garg's result.

The structure of the paper is as follows. Section \ref{LP} introduces a linear programming formulation of the $k$-MST problem. Section \ref{PD} describes the primal-dual subroutine used by Garg, and Section 4 provides an overview of the entire algorithm along with a lower bound on the cost of the optimal tree. In Sections \ref{lambda} and \ref{picking and pruning}, we describe the details of finding a specific tree through parameter setting and modifying the results of the primal-dual subroutine. Finally, Section \ref{approx} proves the 2-approximation.

\section{Linear Programming Formulation} \label{LP}
In this section, we provide a linear programming (LP) relaxation of the $k$-MST problem. The constraints in the dual LP will determine when an $event$ happens in the primal-dual subroutine.

For each $S\subseteq V$, let variable $z_S\in\{0,1\}$ denote whether $S$ constitutes the vertices of the spanning tree. For each edge $e\in E$, let variable $x_e\in\{0,1\}$ denote whether edge $e$ is included in the spanning tree. Then, the following is a linear programming relaxation for the k-MST problem:

\begin{align*}
\text{minimize}     &~~     \sum_{e\in E} c_e x_e \\ 
\text{subject to}   &       \sum_{e:e\in\delta(S)} x_e \geq \sum_{S_T:S\subset S_T} z_{S_T} \qquad \forall S\subset V, \\
                    &~~     \sum_{S\subseteq V} |S|z_S \geq k, \\
                    &~~     \sum_{S\subseteq V} z_S \leq 1, \\
                    &~~~     z_S,~x_e \geq 0.
\end{align*}

The first constraint guarantees that the spanning tree is connected. If $S$ is a strict subset of the vertices $S_T$ of a spanning tree $T$, then there must be at least one edge across $\delta(S)$, the cut of $S$. Given there are no negative-cost edges, any optimal integral solution to the LP will be a tree and have $x_e \leq 1$ for each $e \in E$, and thus these constraints are omitted. We shall also be careful with our subsequent approximation algorithm to not violate these two constraints. We can now write down the dual of this linear program:

\begin{align*}
\text{maximize}     &~~     \lambda_1 k - \lambda_2 \\
\text{subject to}   &       \sum_{S:e\in\delta(S)} y_S \leq c_e \qquad \forall e\in E,\\
                    &~~     \sum_{T\subset S} y_T + \lambda_2 \geq \lambda_1 |S| \qquad \forall S\subset V,\\
                    &~~     \lambda_1,\lambda_2,y_S \geq 0.
\end{align*}

To produce a tree, we use a primal-dual subroutine for these formulations that will be described in the next section. However, we first observe (similar to Paul et al.) that for any $\lambda_1$ and $y$ satisfying the edge constraints, we can find a $\lambda_2$ value satisfying the subset constraints. Particularly, we can let $\lambda_2$ be the maximum of 0 and $\max_{S \subset V}\{\lambda_1 |S| - \sum_{S_T \subset S} y_{S_T}\}$ and we have a feasible dual solution. Therefore, a key component of this algorithm and the following analysis is choosing a $\lambda_1$ value leading to a primal-dual subroutine solution with a tree of the appropriate size. In what follows, we'll see that too small of a $\lambda_1$ value leads to too few selected edges, whereas a value of $\lambda_1$ that is too large leads to  too many selected edges. More details are provided in Section \ref{lambda}.

\section{Primal-Dual Subroutine} \label{PD}

We now present the primal-dual subroutine used by Garg. The algorithm assumes a fixed $\lambda_1$ and, instead of explicitly finding the minimal $\lambda_2$ value, greedily grows a forest with respect to a heuristic function of the dual variables. In our case, the function is the \textit{potential} that we define as below.

\begin{definition}
For any subset $S\subseteq V$, the \textbf{potential} of $S$ is
    $$\pi(S) = \lambda_1|S|-\sum_{T:T\subset S}y_T.$$
\end{definition}

\begin{definition}
A subset $S\subseteq V$ is \textbf{neutral} if $\displaystyle \sum_{T:T\subseteq S}y_T=\lambda_1|S|$, or equivalently, if $y_S=\pi(S)$.
\end{definition}

Intuitively, each vertex has a budget of $\lambda_1$, and we spend the budget amassed in connecting vertices to cover the costs of the edges. In this way, we are able to consider the vertices and edges of a set $S$ as a single variable $\pi(S)$ that measures how much budget we have left to spend on future edges. The objective is to connect $k$ vertices with the cheapest possible edge cost; this corresponds well to maximizing the potential. 

\begin{algorithm}
\caption{Primal-Dual Subroutine PD($\lambda_1$)}
\label{alg:pd}
\begin{algorithmic}
\STATE $y_S \gets 0$
\STATE PD $\gets \emptyset$
\STATE $\mathcal{C} \gets \{\{v\}:v\in V\}$
\WHILE{$\mathcal{C}\neq\emptyset$}
    \STATE raise all $y_S$ corresponding to active components uniformly until either
    \IF{$y_S=\pi(S)$}
        \STATE $\mathcal{C} \gets \mathcal{C}-S$
    \ELSIF{$\sum_{S:e\in\delta(S)}y_S=c_e$ for some $e$ between sets $S_1, S_2$}
        \STATE PD $\gets$ PD $\cup\{e\}$
        \STATE $\mathcal{C} \gets \mathcal{C}-S_1-S_2$
        \STATE $\mathcal{C} \gets \mathcal{C}\cup\{S_1\cup S_2\}$
    \ENDIF
\ENDWHILE
\RETURN PD
\end{algorithmic}
\end{algorithm}

The primal-dual subroutine is given in Algorithm \ref{alg:pd}. Described in words, initially $y_S=0$ for all $S$, and all sets consisting of a single vertex are active. At any stage of the algorithm with active components, we uniformly increase $y_S$ corresponding to all active components until either a set event or an edge event happens. Here, a \textit{set event} is a set becoming neutral, while an \textit{edge event} is the constraint corresponding to an edge becoming \textit{tight} (that is, the dual constraint is met with equality). If a set becomes neutral, then we mark this set inactive and remove it from the set of active components. If the dual constraint for an edge between sets $S_1$ and $S_2$ reaches equality, then the edge is tight. We add this edge to the set of selected edges, mark $S_1$ and $S_2$ inactive if not already, and mark $S_1\cup S_2$ active.  We'll sometimes say that we have {\em merged} the two sets.

We make a few observations about the structure of the result of the primal-dual subroutine. First, the collection $\mathcal{S}$ of all sets that are ever active during the subroutine is \textit{laminar}; that is, for any pair of sets $A, B \in \mathcal{S}$, either $A \subseteq B$, $B \subseteq A$, or $A \cap B = \emptyset$. Initially, all sets consisting of a single vertex are active. By design, the only way a new, larger set becomes active is through merging two previously active sets when an edge event occurs. This maintains the laminar property. Secondly, the subroutine returns a forest. Each time an edge event occurs, two trees are connected into a larger tree. Furthermore, no cycles can exist. If an edge $(u,v)$ was added that created a cycle, there must have been an active set that contained exactly one of $u$ or $v$. However, since $u$ and $v$ must already be connected (otherwise this edge would not complete a cycle), any active set that contains $u$ or $v$ must contain them both. If two edges that would complete a cycle go tight at the same time, we choose one of the edges through a tie-breaking procedure described in Sections \ref{lambda} and \ref{picking and pruning}.

We observe that this procedure always maintains a feasible dual solution. Note that all dual edge constraints are initially satisfied with $y_S = 0$. Furthermore, if an edge becomes tight, the sets corresponding to the endpoints of the edge are marked inactive, so the constraint will never be violated.



Active sets play a crucial role in the subroutine, as these are the sets with the capability for growth. Because of this, we formalize the decomposition of currently-active or once-active sets into neutral subsets and subsets of always-active vertices through the notion of a \textit{kernel}, one of the key points of distinction between our work and Garg's initial presentation as well as Paul et al.'s work. 

\begin{definition}
For an active set $S$ corresponding to a tree $T$ in the set of tight edges of $PD$, the \textbf{kernel} of $S$, denoted by $K(S)$, is the smallest cardinality subset of $S$ such that
\begin{itemize}
    \item[1.] if $v \in S$ has always been part of an active set, then $v \in K(S)$,
    \item[2.] $K(S)$ is connected in $T$, and
    \item[3.]for every once-active set $I \subset S$ either $I \subset K(S)$ or $I \cap K(S) = \emptyset$.
\end{itemize}

The kernel of a once-active set $S$ is the kernel at the moment $S$ becomes neutral.
\end{definition}

Since initially $y_S=0$ for all subsets $S$ and all sets consisting of a single vertex are active, every vertex is the kernel of itself at the start of the primal-dual subroutine. By the definition of the kernel, we know that the kernel of a once-active (but now inactive) set is the kernel of the set when it went neutral. It remains to understand how the kernel changes as active sets grow throughout the primal-dual subroutine. We illustrate this growth through two possible cases:
\begin{itemize}
    \item[1.] an active set $S$ merges with an inactive set $I$ or
    \item [2.] an active set $S$ merges with another active set $A$.
\end{itemize}

In the first case, the kernel of $S \cup I$ is simply the kernel of $S$. Since $I$ was inactive at the time of the merge, every vertex in $I$ has been part of an inactive set. Therefore adding any vertices of $I$ to $K(S)$ would simply increase the cardinality of the kernel. Since we want the minimal cardinality set maintaining connectivity (and $K(S)$ is already connnected since it was previously a kernel), we do not add any of the vertices of $I$. See Figure 1.


\begin{figure}
    \centering
    \begin{subfigure}{.45\textwidth}
    \centering
    \includegraphics[width=6cm]{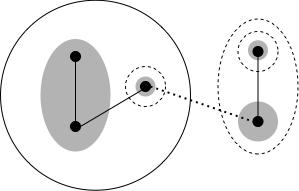}
    \caption{An active and inactive set with their kernels before an edge event occurring at the dashed edge.}
    \end{subfigure}
    \begin{subfigure}{.45\textwidth}
    \centering
    \includegraphics[width=6cm]{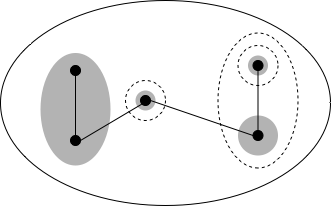}
    \caption{The new active set and its kernel after the merge.}
    \end{subfigure}
    \caption{These figures illustrate how the sets and kernels change when an active set merges with an inactive set. Sets surrounded by a bold (resp. dashed) line are active (resp. inactive). Sets with a grey background are the kernel of the smallest active or inactive set they are contained in. }
    \label{fig:kernel}
\end{figure}

In the second case, the kernel of $S \cup A$ will contain both the kernel of $S$ and the kernel of $A$ since both of these contain vertices which have always been contained in active sets. However, we must be careful since the edge event merging $S$ and $A$ may not occur on an edge connecting the kernel of $S$ to the kernel of $A$ and the resulting set should be connected. In this case, we must add the fewest possible number of once-active subsets of $S$ and $A$ by including only the ones on the path from $K(S)$ to $K(A)$. This maintains that $K(S \cup A)$ is connected in the new tree. See Figure 2.


\begin{figure}
    \centering
    \begin{subfigure}{.45\textwidth}
    \centering
    \includegraphics[width=6cm]{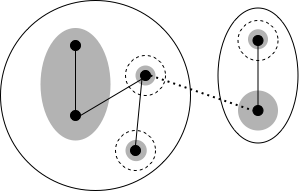}
    \caption{Two active sets with their kernels before an edge event occurring at the dashed edge.}
    \end{subfigure}
    \begin{subfigure}{.45\textwidth}
    \centering
    \includegraphics[width=6cm]{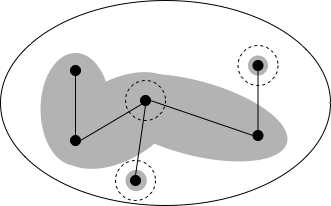}
    \caption{The new active set and its kernel after the merge.}
    \end{subfigure}
    \caption{These figures illustrate how the sets and kernels change when an two active sets merge. Sets surrounded by a bold (resp. dashed) line are active (resp. inactive). Sets with a grey background are the kernel of the smallest active or inactive set they are contained in. }
    \label{fig:kernel2}
\end{figure}

The kernel is used in the next step of the algorithm after the primal-dual subroutine -- the \textit{pruning} phase. In this phase, we want to remove sets of vertices that do not help us achieve our end goal. Particularly, we want to remove neutral sets without disconnecting any tree in the forest returned by the primal-dual subroutine. We do so by pruning each tree $T$, with vertex set $S_T$, in the forest returned by $PD$ to return exactly the edge-set determined by the kernel of $S_T$. In particular, a pruned tree $T$ is given by $E(K(S_T))\cap T$ where $E(S) = \{(u,v) \in E: u,v \in S\}$. The following lemma shows that this has our desired effect; the proof is delayed until Section \ref{approx}. For sets $S' \subset S \subset V$, let $\delta_{K(S)}(S') = \{(u,v) \in E(S) : u \in S', v \in K(S) - S'\}.$  

\begin{lemma}\label{lemma:prune}
The kernel $K(S)$ of a set $S$ contains no neutral subset $N \subset K(S)$ such that
$|\delta_{K(S)}(N)| = 1.$
\end{lemma}

Clearly the size of each tree may decrease during the pruning phase, so measures must be taken to ensure we have a tree of suitable size at the end of the primal-dual subroutine.
Recalling that we can find a feasible $\lambda_2$ value for any $\lambda_1$ and feasible $y_S$ solution, we carefully select $\lambda_1$ guaranteeing at least one kernel in our primal-dual output forest has at least $k$ vertices. After pruning, we execute a picking routine to obtain a tree with exactly $k$ vertices such that the 2-approximation holds.  The details of this selection of $\lambda_1$ and the picking routine implemented after the primal-dual subroutine and pruning occur are described in the following sections.


\section{Algorithm Overview}

In this section, we provide an overview of how the forest from the primal-dual subroutine is used to construct a feasible $k$-MST. Assume we have run the primal-dual subroutine with some value of $\lambda_1$ to find a feasible dual solution $(y, \lambda_1, \lambda_2)$, though we may not know $\lambda_2$ exactly. We also have a forest $F$ from which we need to select a tree spanning $k$ vertices. In order to do so, we need to guarantee that there exists a tree in $F$ containing at least $k$ vertices after it is pruned. If the fixed $\lambda_1$ is too small, this may not be the case as a small $\lambda_1$ allows sets to become neutral earlier which limits the potential for growth. On the other hand, if $\lambda_1$ is too large, the primal-dual subroutine degrades to greedily selecting the cheapest edges, which is sub-optimal. In order to balance these two scenarios, we search for an appropriate $\lambda_1$ value. Specifically, we identify a $\lambda_1$ value such that $PD(\lambda_1^-)$ returns a forest where every pruned tree is too small and $PD(\lambda_1^+)$ contains at least one pruned tree large enough. Here $x^-=x-\epsilon$ and $x^+=x+\epsilon$, where $\epsilon$ is arbitrarily small. Once we have this threshold value of $\lambda_1$, we prune our selected tree to return a kernel with at least $k$ vertices and bounded cost on the corresponding tree. Finally, we select a sub-tree of our pruned tree containing exactly $k$ vertices.

Before describing how we choose a $\lambda_1$ value, we derive a lower bound of the cost of the optimal $k$-MST in terms of the potential function that works for any choice of $\lambda_1$. In the following sections, we will see how to set $\lambda_1$ to give an upper bound. Together, this will give us a 2-approximation. 

Let $\mathcal{S}$ be the laminar set of all once-active sets plus the set of all vertices. Denote $T^*$ the optimal $k$-MST, $S_{T^*}$ its set of vertices, and $S_1$ the set with the minimal potential in $\mathcal{S}$ such that $S_{T^*} \subset S_1$. Since $V \in\mathcal{S}$, such an $S_1$ always exists. In general, for a tree, we will refer to the set of edges by $T$ and the corresponding set of vertices by $S_T$.

\begin{lemma}
For any $S\subseteq V$, $\displaystyle \sum_{U:U\subseteq S}y_U\leq\lambda_1|S|$.
\end{lemma}

\begin{proof}
If $S$ was once active, then the inequality holds by the design of the primal-dual subroutine. We prove this by induction. Suppose $S_1$ and $S_2$ merged to form $S$, and 
$$\sum_{U:U\subseteq S_1}y_U\leq\lambda_1|S_1|,$$
$$\sum_{U:U\subseteq S_2}y_U\leq\lambda_1|S_2|.$$
Initially $y_S=0$, so
\begin{align*}
    \sum_{U:U\subseteq S}y_U 
        &= \sum_{U:U\subseteq S_1}y_U + \sum_{U:U\subseteq S_2}y_U + y_S \\
        &\leq \lambda_1|S_1| + \lambda_1|S_2| \\
        &= \lambda_1|S|.
\end{align*}
We increase $y_S$ either until $S$ merges with another active set or until $\displaystyle \sum_{U:U\subseteq S}y_U = \lambda_1|S|$ and $S$ gets marked neutral. In either case, we no longer increase $y_S$, so the claim continues to hold. For an arbitrary set $S$, we can partition it into maximal disjoint laminar subsets $S_1,S_2,...,S_c \in \mathcal{S}$. Therefore,
$$\sum_{U:U\subseteq S}y_U = \sum_{i=1}^{c} \sum_{U:U\subseteq S_i} y_U
\leq \sum_{i=1}^c \lambda_1|S_i| = \lambda_1|S|.$$
\end{proof}

\begin{theorem}
The minimal spanning tree has cost at least $\lambda_1 \cdot k-\pi(S_1)$.
\end{theorem}

\begin{proof}
By the potential definition and Lemma 1, 
\begin{align*}
    \lambda_1|S_1| &= \sum_{U:U\subset S_1}y_U + \pi(S_1) \\
                   &= \sum_{U:U\subseteq S_1-S_{T^*}}y_U + \sum_{\substack{U:U\subset S_1 \\ U\cap        S_{T^*} \neq \emptyset}}y_U + \pi(S_1) \\
                   &\leq \lambda_1|S_1-S_{T^*}| + \sum_{\substack{U:U\subset S_1 \\ U\cap S_{T^*} \neq        \emptyset}}y_U + \pi(S_1), \text{ so} \\
    \lambda_1|S_{T^*}| &\leq \sum_{\substack{U:U\subset S_1 \\ U\cap S_{T^*} \neq \emptyset}}y_U + \pi(S_1) \\
                   &\leq \sum_{e\in S_{T^*}}\sum_{U:e\in\delta(U)} y_U + \pi(S_1) \\
                   &\leq \sum_{e\in S_{T^*}} c_e +  \pi(S_1). 
\end{align*}
The first inequality follows from Lemma 1. The third inequality says that for every set that intersects $S_{T^*}$, an edge in its cut must lie in $T^*$. Since we only allow edges with non-negative costs, the spanning tree is minimal when it covers exactly $k$ vertices. By rearranging, we obtain the claim.
\end{proof}

\section{Setting \texorpdfstring{$\lambda_1$}{parameters}} \label{lambda}


\begin{figure}[]
    \centering
    \includegraphics[width=7cm]{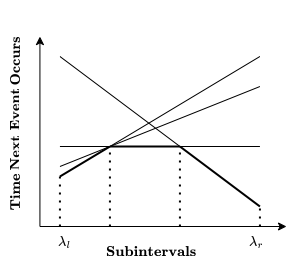}
    \caption{\textbf{Finding Threshold $\lambda_1$.} Each line represents the time for an event to occur next in terms of $\lambda_1$. The bold line shows the next event, where each segment is from a given subinterval.}
\end{figure}

We now describe how to set $\lambda_1$ to enable us to pick $k$ vertices from a pruned tree in our primal-dual subroutine forest. Recall  $x^-=x-\epsilon$ and $x^+=x+\epsilon$, where $\epsilon$ is arbitrary small. We will later use \textit{infinitesimal} to refer to variables that approximate their originals as $\epsilon \rightarrow 0$. We adapt two lemmas from Paul et al.\ to the $k$-MST situation. The first tells us that we can find our desired threshold value $\lambda_1$.

\begin{lemma}
In polynomial time, we can find a threshold value $\lambda_1$ such that all pruned trees of PD($\lambda_1^-$) have less than $k$ vertices and there exists at least one pruned tree in PD($\lambda_1^+$) with at least $k$ vertices.
\end{lemma}

As the details of this proof closely mirror those of Paul et al., we omit them here and refer the reader to Lemma 2 \cite{PaulFFSW20}. The key idea is that the time for each set and edge event to occur can be represented as a linear function in $\lambda_1$; see Figure 3. Then by observing the threshold $\lambda_1$ must occur at an intersection of these lines, we can consider smaller and smaller intervals of $\lambda$ determined by the intersections until we find our desired value. In particular, we recurse onto a smaller subinterval where the next event to occur is consistent throughout the subinterval and consider the possible events that can follow. Eventually, there must be a time (intersection) where the difference in next events translates to a difference in sizes of the resulting kernels; this is where our threshold $\lambda_1$ occurs. The next lemma tells us two important properties regarding the primal-dual subroutine for values around our threshold $\lambda_1$.

\begin{lemma}
Throughout the two subroutines $PD(\lambda_1^-)$ and $PD(\lambda_1^+)$, the following two properties hold: 
\begin{itemize}[itemsep=0mm]
    \item All active components are the same except for during infinitesimal time.
    \item For all $S \subseteq V$, the difference between $y_S^+$ and $y_S^-$ is infinitesimal. Here $y_S^+$ and $y_S^-$ are the dual variables when running $PD(\lambda_1^+)$ and $PD(\lambda_1^-)$, respectively.
\end{itemize}
\end{lemma}

Here, again we present an overview of the proof of Lemma 3 \cite{PaulFFSW20} with a few minor changes. The main observation is the claim could only fail if two different events occur at the same time in $PD(\lambda)$, and the events cause lasting disparity in $PD(\lambda_1^-)$ and $PD(\lambda_1^+)$. Furthermore, there are four possible ways that two different events could occur at the same time:
\begin{itemize}
    \item[1.] different sets go neutral in $PD(\lambda_1^-)$ and $PD(\lambda_1^+)$,
    \item[2.] an edge goes tight in $PD(\lambda_1^-)$ while a set goes neutral in $PD(\lambda_1^+)$,
    \item[3.] a set goes neutral in $PD(\lambda_1^-)$ while an edge goes tight in $PD(\lambda_1^+)$, or
    \item[4.] different edges go tight in $PD(\lambda_1^-)$ and $PD(\lambda_1^+)$.
\end{itemize}

In the first case, the times for the two sets to go neutral must differ by an infinitesimally small amount and thus one will go neutral immediately after the other. The second case cannot occur for this problem since the time for a set to go neutral has a positive slope in $\lambda_1$ while the time for an edge to go tight has a negative slope in $\lambda_1$. 

For the third case, if the tight edge has an endpoint in an active set different than the set going neutral, the edge will still go tight immediately after the set event. If the edge going tight in $PD(\lambda_1^+)$ merges the set going neutral in $PD(\lambda_1^-)$ and another neutral set, the merged set must have infinitesimally small potential and will go neutral immediately after the edge event. This maintains the active sets. Furthermore, if an active set merges with the set going neutral in the future in $PD(\lambda_1^-)$, the edge will go tight immediately after, still maintaining the active sets.

Finally, for the fourth case, if the edges are between different components one will go tight immediately after the other (similar to the first case). Meanwhile, if they are between the same components, only one will go tight in each subroutine but the resulting active components will be the same.

The main role of Lemma 4 is insight into the potential differences between $PD(\lambda_1^-)$ and $PD(\lambda_1^+)$. In particular, there may be subsets marked neutral in $PD(\lambda_1^-)$ but with infinitesimally small potential in $PD(\lambda_1^+)$ or there may be different edges that went tight between the same components. If, every time two events tied in $PD(\lambda)$, we broke the tie by selecting what $PD(\lambda_1^+)$ would do, we will end up with at least one kernel having at least $k$ vertices. On the other hand, we can consider breaking ties in favor of $PD(\lambda_1^-)$ one at a time. By doing so, we will find the smallest $i$ such that if the first $i$ ties are broken according to $PD(\lambda_1^-)$ and the rest by $PD(\lambda_1^+),$ we return a forest with all kernels containing less than $k$ vertices. By Lemma 3, the dual variables only change by an infinitesimally small amount, and the only differences occur during the pruning phase.

In finding this value of $i$, we have also either identified a neutral subset $X$ such that if $X$ remained active our forest would contain a kernel of appropriate size, or found two edges $e$ and $f$ between the same components such that adding $f$ instead of $e$ returns a forest containing a kernel of appropriate size. These two cases will play a role in picking our final set of vertices in the next section.

\section {Constructing a Tree}\label{picking and pruning}

Let $\lambda_1$ be our found threshold value and $(y, \lambda_1, \lambda_2)$ our feasible dual solution acquired through tie-breaking in the manner described at the end of Section 5. Currently, all kernels of our primal-dual subroutine output forest have fewer than $k$ vertices, but Section 5 tells us that either we have identified a neutral subset $X$ such that if $X$ remained active our forest would contain a kernel with at least $k$ vertices (Case I) or found two edges $e$ and $f$ between the same components such that adding $f$ instead of $e$ returns a forest containing a kernel with at least $k$ vertices (Case II). The final construction of our tree depends on which of these cases are present and requires us to pick $k$ vertices. Specifically, \textit{pick}$(X,w,k)$ returns a sub-tree of $X$ with $k$ vertices and contains the vertex $w$. The idea is to inspect the last two subsets that merged to form the set. Suppose that $X_1$ and $X_2$ merged to form $X$, with edge $(u,v)$ connecting them, and that $X_1$ contains $w$. If $X_1$ contains at least $k$ vertices, then we invoke \textit{pick}$(X_1,w,k)$. If $X_1$ has less than $k$ vertices, then we pick all vertices in $X_1$ and continue to \textit{pick}$(X_2,v,k-|X_1|)$. We repeat the process recursively until we have picked exactly $k$ vertices. 

\begin{algorithm}
\caption{Pick Routine \textit{pick}$(X,w,k)$}
\begin{algorithmic}
\STATE let $X_1$ and $X_2$ be the two subsets that merged on edge $e=(u,v)$ to form $X$
\STATE suppose without loss of generality that $w \in X_1$
\IF{$|X_1| > k$} 
    \STATE call \textit{pick}($X_1, w, k$)
\ELSIF {$|X_1| < k$} 
    \STATE pick all vertices in $X_1$
    \STATE call \textit{pick}($X_2, v, k-|X_1|$)
\ELSE
    \STATE pick all vertices in $X_1$
\ENDIF
\end{algorithmic}
\end{algorithm}

In Case I, a set goes neutral the same time an edge goes tight; see Figure 4. Let $K(S_1)$ and $K(S_2)$ be the two kernels of the merging two subsets $S_1$ and $S_2$. If we break the tie by choosing the set event, then we would end up with both $K(S_1)$ and $K(S_2)$ having less than $k$ vertices. On the other hand, if we break the tie by choosing the edge event, then the new kernel $K(S_1\cup S_2) = K(S_1) \cup N_1 \cup \dots \cup N_p \cup K(S_2)$ would have at least $k$ vertices. Here $N_i$ denotes the neutral sets on the path from $K(S_1)$ to $K(S_2)$. Now we show how to pick exactly $k$ vertices from this new kernel using the pick routine. Starting from $K(S_1)$, we select neutral sets $N_1, N_2, \dots, N_{q-1}$ until adding another neutral set $N_q$ will cause  $K(S_1) \cup N_1 \cup \dots \cup N_q$ to have at least $k$ vertices. Suppose edge $e=(u,v)$ links $N_{q-1}$ to $N_q$, then \textit{pick}$(N_q, v, r)$ will pick the remaining vertices needed, where $r = k-|K(S_1)|-\sum_{i=1}^{q-1}|N_i|$ is the number of additional vertices we need to pick.

\begin{figure}[]
    \centering
    \begin{subfigure}{\textwidth}
    \centering
    \includegraphics[width=10.5cm, height=4.5cm]{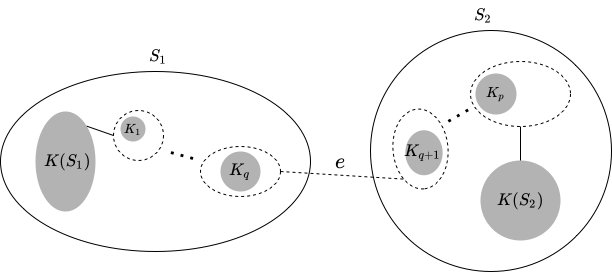}
    \caption{Sets $S_1$ and $S_2$ and their kernels before the edge event causing $e$ to go tight and the set event causing $S_2$ to go neutral tie in $PD(\lambda_1)$.}
    \end{subfigure}
    \begin{subfigure}{\textwidth}
    \centering
    \includegraphics[width=10.5cm, height=4.5cm]{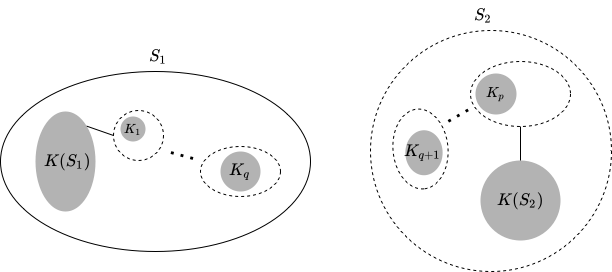}
    \caption{Sets $S_1$ and $S_2$ and their kernels after the tie is broken by allowing $S_2$ to go neutral.}
    \end{subfigure}
    \begin{subfigure}{\textwidth}
    \centering
    \includegraphics[width=10.5cm, height=4.5cm]{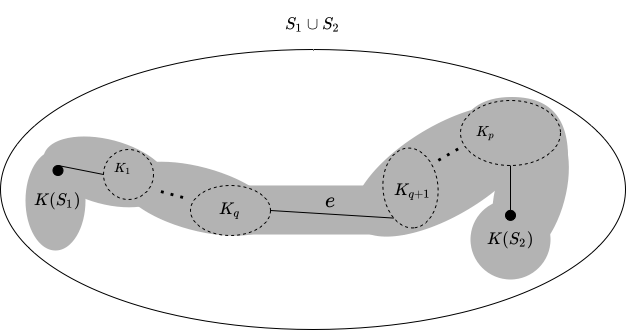}
    \caption{Set $S_1 \cup S_2$ and its kernel after the tie is broken by allowing $e$ to go tight.}
    \end{subfigure}
    \caption{\textbf{Case I.} $S_2$ goes neutral the same time $e$ goes tight.}
\end{figure}

In Case II, two edges go tight simultaneously; see Figure 5. Again, let $K(S_1)$ and $K(S_2)$ be the two kernels of the merging two subsets $S_1, S_2$. Denote $e, f$ the two edges both between $S_1$ and $S_2$. If we choose edge $e$, then the new kernel $K(S_1\cup S_2) = K(S_1) \cup N_1 \cup ... \cup N_p \cup K(S_2)$ would have less than $k$ vertices, where $N_1, \dots, N_p$ are the neutral sets between $K(S_1)$ and $K(S_2)$ using edge $e$. On the other hand, if we choose edge $f$, then the new kernel $K(S_1\cup S_2) = K(S_1) \cup N'_1 \cup ... \cup N'_q \cup K(S_2)$ would have at least $k$ vertices, where $N'_1, \dots, N'_q$ are the neutral sets between $K(S_1)$ and $K(S_2)$ using edge $f$. The difference clearly results from the neutral sets in between $K(S_1)$ and $K(S_2)$. Once again, we pick $K(S_1), N'_1, N'_2, \dots, N'_{t-1}$ where picking $N'_t$ would give at least $k$ vertices and invoke \textit{pick}$(N'_t, v, r)$ to finish the rest ($N'_t,v,r$ defined similar to Case I).

\begin{figure}[]
    \centering
    \begin{subfigure}{\textwidth}
    \centering
     \includegraphics[width=8cm, height=4.5cm]{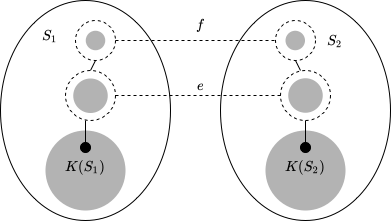}
     \caption{Sets $S_1$ and $S_2$ and their kernels before the edge events causing $e$ and $f$ to go tight tie in $PD(\lambda_1)$.}
    \end{subfigure}
    \begin{subfigure}{.45\textwidth}
    \centering
     \includegraphics[width=4cm]{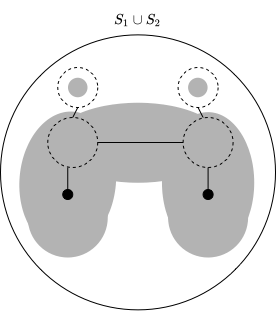}
     \caption{Set $S_1 \cup S_2$ and its kernel after the tie is broken so $e$ goes tight.}
    \end{subfigure}%
    \begin{subfigure}{.45\textwidth}
    \centering
     \includegraphics[width=4cm]{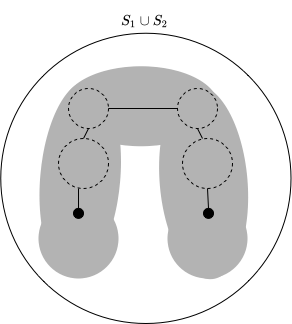}
     \caption{Set $S_1 \cup S_2$ and its kernel after the tie is broken so $f$ goes tight.}
    \end{subfigure}
   
    \caption{\textbf{Case II.} Edges $e$ and $f$ between $S_1$ and $S_2$ go tight at the same time.}
\end{figure}


\section{2-Approximation}\label{approx}

Now that we have finished describing the mechanics of the algorithm, we can finally present the proof of the 2-approximation. Let $T'$ denote the tree we obtained after pruning with vertex set (kernel) $K'$, $T_0\subseteq K'$ the tree of $k$ vertices we obtained by the pick routine, and $S_{T_0}$ the set of vertices of $T_0$. Further, let $v$ be the vertex in $S_{T_0}$ where the pick routine terminated.
Before proceeding with the necessary lemmas, we present the proof of Lemma \ref{lemma:prune} from Section \ref{PD}.

\begin{proof}
We prove this by induction on the number of events. In the base case, at the start of the primal-dual subroutine, every vertex is the kernel of itself and no neutral sets have appeared yet. 

Suppose that for the first $k$ events the claim holds, and we consider the next event. If the $(k+1)$-st event is a set event, then a set $S$ goes neutral. A set going neutral does not affect its kernel, so as previously, the claim still holds for $S$; other sets are also unaffected so the claim holds for them. 

If the $(k+1)$-st event is an edge event, then an edge $e$ between sets $S_1$ and $S_2$ goes tight and the two sets merge to form a new active set $S$. First of all, this event does not affect the kernel of $S_1$, $S_2$, or of any other set besides $S$, so the claim still holds for all of them. Since at least one of the merging sets is active, we can suppose without loss of generality that $S_1$ is active. Then, if $S_2$ is inactive, the kernel of $S$ would be the kernel of $S_1$. For a neutral set $N \subset K(S)$, $|\delta_{K(S)}(N)| = 1$ for some $N$ would imply that $|\delta_{K(S_1)}(N)| = 1$, which contradicts that the claim holds previously.

If $S_2$ is active, then the kernel of $S$ would be the kernel of $S_1$ and $S_2$, plus some neutral sets on the path between the two kernels. If $N \subset K(S)$, then either $N$ is on the path between $K(S_1)$ and $K(S_2)$ or $N$ is a subset of $K(S_1)$ or $K(S_2)$. But every neutral set $N$ on the path has $|\delta_{K(S)}(N)| = 2$, and by the inductive hypothesis, no neutral subset $N$ of $K(S_1)$ or $K(S_2)$ can have $|\delta_{K(S_i)}(N)| = 1$, $i=1,2$.

\end{proof}

Again, we utilize a result by Paul et al.\ (Lemma 4 in \cite{PaulFFSW20}).

\begin{lemma}\label{lemma:dualbound}
$$\displaystyle \sum_{e\in T_0}\sum_{S:e\in \delta(S)}y_S \leq 2\sum_{\substack{U:U\cap S_{T_0}\neq\emptyset \\ v\not\in U}}y_U.$$
\end{lemma}

\begin{proof}
We will prove the inequality for any arbitrary iteration in the primal-dual algorithm. Consider an iteration in which we let $\mathcal{C}$ be the current set of components $C$ such that $|\delta(C) \cap T_0| \geq 1$. We can partition $\mathcal{C}$ into active components $\mathcal{C}_A$ and inactive components $\mathcal{C}_I$. Let $v$ be the final vertex picked and $C_v$ be the unique set in ${\cal C}$ containing $v$. 

We claim that if $C_v \in {\cal C}_A$, then $\sum_{C \in {\cal C}_A} |\delta(C) \cap T_0| \leq 2|{\cal C}_A|-2$, otherwise if $C_v \in {\cal C}_I$, then $\sum_{C \in {\cal C}_A} |\delta(C) \cap T_0| \leq 2|{\cal C}_A|-1$.  Suppose for now that the claim is true.  We now prove the inequality in Lemma 4 by induction on the algorithm. At the start of the algorithm, with $y_S = 0$ for all $S$,  both sides of the inequality are equal to $0$. At each iteration, let $\epsilon$ be the amount that we raise $y_{C}$ for each active component $C \in \mathcal{C}_A$. The LHS of the inequality increases by $ \sum_{C\in \mathcal{C}_A} |\delta(C)\cap T_0|\epsilon$, while the RHS of the inequality increases by either $2|{\cal C}_A|\epsilon$ (if $C_v \in {\cal C}_I$) or $2(|{\cal C}_A|-1)\epsilon = (2|{\cal C}_A| - 2)\epsilon$ (if $C_v \in {\cal C}_A$).  Then given the claim, the inequality continues to hold inductively.  Thus the lemma statement will hold at the end of the algorithm.

To prove the claim, first suppose that $C_v$ is inactive. By Lemma 1, all other neutral subsets of $S_{T_0}$ have degree at least $2$. Since $v$ is the last vertex added, $C_v$ is the only inactive component such that possibly $|\delta(C_v) \cap T_0| = 1$. Thus we have
$$\sum_{C\in\mathcal{C}_I} |\delta(C) \cap T_0| \geq 2|\mathcal{C}_I| - 1.$$

Note that edges of $T_0$ link components in $\mathcal{C}$ to form a tree, so
$$\sum_{C\in\mathcal{C}_A} |\delta(C)\cap T_0| + \sum_{C\in\mathcal{C}_I} |\delta(C)\cap T_0| \leq 2|\mathcal{C}_A| + 2|\mathcal{C}_I| - 2.$$ Then the last two inequalities imply
$$\sum_{C\in\mathcal{C}_A} |\delta(C)\cap T_0| \leq 2|\mathcal{C}_A| - 1,$$ and the claim holds for this case.

Now consider the case where $C_v \in \mathcal{C}_A$. By a similar logic, there is no component $C \in \mathcal{C}_I$ such that $|\delta(C) \cap T_0| = 1$, and therefore
$$\sum_{C\in\mathcal{C}_I} |\delta(C) \cap T_0| \geq 2|\mathcal{C}_I|, \text{ implying}$$
$$\sum_{C \in \mathcal{C}_A} |\delta(C)\cap T_0| \leq 2|\mathcal{C}_A| - 2,$$ and the claim  holds for this case, so the proof of the lemma is complete.
\end{proof}

The result of Lemma \ref{lemma:dualbound} allows us to prove the following upper bound on the cost of our tree.

\begin{theorem}
The picked tree has cost at most $2(\lambda_1 \cdot k - \pi(S_2))$, where $S_2$ is the maximal potential set that contains the picked tree.
\end{theorem}

\begin{proof}
By the pick procedure, vertices in $S_2-S_{T_0}$ are either (i) in a pruned neutral subset $N'_i$ or (ii) in the subset $N_p$ where we started our pick procedure. Thus, we have $S_2=\bigcup N'_i\cup(N_p-S_{T_0})\cup S_{T_0}$. Since $N'_i$ are neutral, we have 
$$\displaystyle \lambda_1|\cup N'_i|=\sum_{U:U\subseteq \cup N'_i}y_U.$$

$N_p$ is also neutral, and we can partition its subsets into two types: ones that contain vertices in $N_p - S_{T_0}$ and ones that do not. Then we have 
$$\lambda_1|N_p| = \sum_{\substack{U:U\subset N_p \\ U\cap (N_p - S_{T_0}) \neq \emptyset}}y_U + \sum_{U:U\subseteq N_p\cap S_{T_0}}y_U 
    \leq \sum_{\substack{U:U\subseteq N_p \\ U\cap (N_p- S_{T_0}) \neq \emptyset}}y_U + \lambda_1|N_p\cap S_{T_0}|$$ by Lemma 1 which implies $$\lambda_1|N_p-S_{T_0}| \leq \sum_{\substack{U:U\subseteq N_p \\ U\cap (N_p - S_{T_0}) \neq \emptyset}}y_U.$$

Combining with Lemma \ref{lemma:dualbound}, this gives us
\begin{align*}
    \lambda_1|S_2| &=    \sum_{U:U\subset S_2}y_U + \pi(S_2) \\
                 &\geq \sum_{\substack{U:U\cap S_{T_0}\neq\emptyset \\ v\not\in U}}y_U
                 +     \sum_{\substack{U:U\subset N_p \\ U\cap (N_p - S_{T_0}) \neq \emptyset}}y_U
                 +     \sum_{U:U\subseteq \cup N'_i}y_U + \pi(S_2) \\
                 &\geq \frac{1}{2}\sum_{e\in T_0}\sum_{S:e\in \delta(S)}y_S + \lambda_1|N_p-S_0| + \lambda_1|\bigcup N'_i| + \pi(S_2). 
\end{align*}    
Rearranging gives $\lambda_1|S_{T_0}| \geq \frac{1}{2}\sum_{e\in T_0}c_e + \pi(S_2).$
\end{proof}

Combining our lower bound from Theorem 1 with the upper bound in Theorem 2, we achieve the 2-approximation.

\begin{theorem}
The tree returned by the picked routine has at most twice the cost of the optimal spanning tree of $k$ vertices, that is
$$\sum_{e\in T_0}c_e \leq 2\sum_{e\in T^*}c_e.$$
\end{theorem}

\begin{proof}
Recall from Theorem 1 we have that 
$$\sum_{e\in T^*}c_e \geq \lambda_1 \cdot k-\pi(S_1),$$
where $S_1$ is the set with minimal potential in $\mathcal{S}$ that contains $S_{T^*}$. Since we include the set of all vertices $V(G)$ in $\mathcal{S}$, the fact that both $S_{T^*}$ and $S_{T_0}$ are subsets of $V(G)$ implies that $\pi(S_2) \geq \pi(V(G))$ and $\pi(S_1) \leq \pi(V(G))$. Thus we have
\begin{align*}
    \sum_{e\in T_0} c_e &\leq 2(\lambda_1 \cdot k - \pi(S_2)) &\text{by Theorem 2} \\
                      &\leq 2(\lambda_1 \cdot k - \pi(S_1)) \\
                      &\leq 2\sum_{e\in T^*} c_e.
\end{align*}
\end{proof}

\end{document}